## correspondence

# Tin Pest: A Forgotten Issue in the Field of Applied Superconductivity?

**To the Editor** — In the framework of the hybrid magnet project aiming to produce 43 T in a 34 mm warm aperture at LNCMI-Grenoble, a novel type of superconducting conductor is being developed in collaboration with CEA-Saclay. It is based on the soft soldering of a Nb-Ti flat Rutherford cable on a Cu-Ag hollow stabilizer that will be cooled by superfluid helium *i.e.* a Rutherford Cable on Conduit Conductor (RCOCC)[1]. During the study and the qualification of the soft soldering process, it has been systematically observed that the rupture shear strength of the joint of RCOCC samples prepared with the Sn96Ag4 solder alloy is (i) in the range 20 to 35 MPa at room temperature depending on the bond quality achieved in agreement with published values [2] and, (ii) much lower at 77 K, *i.e.* with a reduction between 30 to 50 % with respect to room temperature values, a somehow opposite behavior to the one usually known and observed for "normal" materials.

To investigate the thermo-mechanical behavior of Sn96Ag4 soldered joints, shear tests have been performed in dedicated samples resulting from the assembly of two copper (Cu) slabs of 99.9 % purity and dimensions of 155x15x2 mm[3]. Twelve test samples have been prepared by joining Cu slabs with a lap joint of $10 \pm 2$ mm length and $15 \pm 0.1$ mm width. Six of them have been made with 0.1 mm thick ribbons of Sn96Ag4 solder alloy covering the contact area and the six other ones with the same quantity of Sn60Pb40. In both cases, copper surfaces have been carefully cleaned with the flux F-SW 21/3.1.1 (Norm DIN 8511/EN 29454). Samples prepared with the Sn96Ag4 soldering alloy (melting point Solidus/Liquidus = 221/221°C) have been heated up to 226°C whereas for Sn60Pb40 ones (melting point Solidus/Liquidus = 183/191°C), the temperature for the soldering has been fixed to 200°C. Mechanical measurements have been performed using a conventional tensile testing machine equipped with a liquid nitrogen reservoir for the tests at 77 K.

Averaged results obtained from shear tests performed on the twelve dedicated samples are summarized in Fig. 1. A clear difference of the thermo-mechanical behavior between Sn96Ag4 and Sn60Pb40 samples tested at 300 K and 77 K can be observed. The rupture shear strength of Sn96Ag4 soldered samples *decreases* in average by 12.9 MPa at low temperature (-37 %) whereas for Sn60Pb40 ones an average *increase* of 8.5 MPa is observed (+28 %). It can be added that the rupture shear force measured at 77 K for Sn96Ag4 soldered samples is in agreement with results recently reported in a CERN internal note[3].

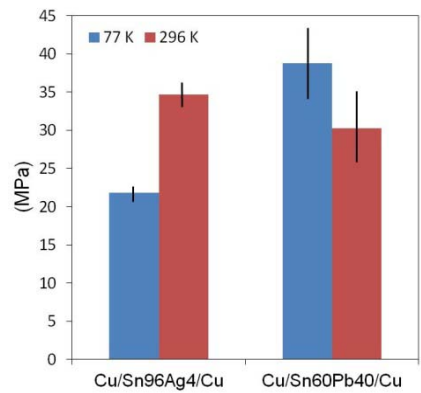

Figure 1 | **Averaged rupture shear strength measured on soldered samples.**

The origin of the difference between the thermo-mechanical behavior of both soldering alloys lies most probably in the tin pest, namely the result of the phase transformation of tin appearing in Sn96Ag4 below 286 K [4,5,6,7]. This allotropic change of the "normal" white β-tin (tetragonal structure, metallic compound) into gray α-tin (diamond like cubic structure, semiconductor compound) is accompanied by a 26 % increase in volume[5] implying serious problems when rupture shear strength or lifetime of a soldered joint is considered. Indeed this transformation favors the apparition of cracks that weaken the mechanical toughness of Sn96Ag4. In Sn60Pb40, lead is known to stabilize the β-tin phase like antimony (Sb) and bismuth (Bi)[6,7] making Sn-Sb and Sn-Bi based alloys possible alternatives for lead-free solders operating below 286 K. But even with a high concentration of lead such as in Sn63Pb37, tin pest can also appear on long term[8] restricting the choice to tin-free solder alloys for high reliability interconnects working at cryogenic temperature.

Even if it is still unproven but not impossible that tin pest has destroyed the buttons of uniforms of Napoleon's army during its disastrous retreat from Russia in 1812, this problem is real and known for a few hundreds of years. It may have become a forgotten issue but experiences currently a strong revival of interest due to the increased environmental and health concerns regarding the toxicity of lead and the possible alternative offered by tin based solders.

In conclusion, more investigations are required to definitely prove the role of tin pest in the weakening of the mechanical toughness of Sn-Ag soldered joints. As an immediate consequence it is strongly recommended to avoid using Sn-Ag binary soldering alloys for joints that will be used at cryogenic temperatures, especially if they are not mechanically re-enforced or if they are submitted to mechanical loading. ❐

References

[1] P. Pugnat, *et al.* to appear in *IEEE Trans. Appl. Supercond.* (2012).

[2] H. Wu, et al. *IEEE Trans. Appl. Supercond.* **21** 1738–1741 (2011).

[3] S. Heck, et al. *CERN-ATS-Note-2011-074*, unpublished (2011);

http://cdsweb.cern.ch/record/1378836/files/ATS_Note_2011_074_ChrScheurlein.pdf

[4] L.D. Brownlee, *Nature* **166**(4220), 482 (1950).

[5] J.H. Becker, *J. Appl. Phys.* **29**(7), 1110 – 1121 (1956).

[6] Y. Kariya, C. Gagg, and W.J. Plumbridge, *Soldering and Surface Mount Technology*, **13**(1) 39–40 (2000); http://www.smartgroup.org/pdf/tinpest.pdf

[7] Y. Kariya, N. Williams, C. Gagg, W. Plumbridge, *Journal of the Minerals, Metals and Materials Society* **53**(6) 39–41 (2001); http://iweb.tms.org/PbF/JOM-0106-39.pdf

[8] W.J. Plumbridge, *Soldering & Surface Mount Technology*, **22**(1) 56–57 (2010).

R. Pfister, and P. Pugnat*
Lab. National des Champs Magnétiques Intenses (LNCMI), CNRS-UJF-UPS-INSA, BP166, 38042 Grenoble Cedex 9, FRANCE
*e-mail: Pierre.Pugnat@lncmi.cnrs.fr